\documentclass[twocolumn,showpacs,amsmath,amssymb,prl,%
               superscriptaddress]{revtex4}
\usepackage[T1]{fontenc}
\usepackage{times}
\usepackage{graphicx,color}
\usepackage{wasysym}
\usepackage{bm}

\begin{document}

\title{Universal Conductance and Conductivity at Critical Points in Integer Quantum Hall Systems} 
\author{L. Schweitzer}
\affiliation{Physikalisch-Technische Bundesanstalt, Bundesallee 100, 38116
Braunschweig, Germany}
\author{P. Marko\v{s}}
\affiliation{Institute of Physics, Slovak Academy of Sciences, 845
11 Bratislava, Slovakia}

\date{\today}

\begin{abstract}
The sample averaged longitudinal two-terminal conductance and the respective 
Kubo-conductivity are calculated at quantum critical points in the integer quantum 
Hall regime. In the limit of large system size, both transport quantities are 
found to be the same within numerical uncertainty in the lowest Landau band, 
$0.60\pm 0.02\,e^2/h$ and $0.58\pm 0.03\,e^2/h$, respectively. 
In the 2nd lowest 
Landau band, a critical conductance $0.61\pm 0.03\,e^2/h$ is obtained which 
indeed supports the notion of universality. However,  these 
numbers are significantly at variance with the hitherto commonly believed  
value $1/2\,e^2/h$. We argue that this difference is due to the
multifractal structure of critical wavefunctions, a property that should 
generically show up in the conductance at quantum critical points.
\end{abstract}

\pacs{72.15.Rn, 71.30.+h}

\maketitle


The transitions between quantized Hall plateaus are generally viewed as
a manifestation of continuous $T=0$ quantum phase transitions. They emerge
in the presence of a strong magnetic field in disordered two-dimensional 
electron or hole gases which can be observed in Silicon MOSFETs and 
GaAs-heterostructures. According to renormalization group considerations, 
some properties of the quantum phase transitions should be universal, i.e., 
independent of microscopic details of the system. For instance, this is 
supposed for the critical exponent that controls the divergence of the 
localization length $\xi(E)=\xi_0 |E-E_c|^{-\mu}$ near the critical energies 
$E_c$, a behavior which has been corroborated by experimental 
\cite{WTPP88,KHKP91} and numerical \cite{CC88,HK90,HB92} studies. 
In addition, it has been suggested that the impurity averaged elements of 
the conductivity tensor act as coupling constants in an appropriate field 
theory \cite{LLP83}
where all the transitions belong to the same 
universality class \cite{FGG90}
with universal values of the critical
dissipative and Hall conductivities. It has been proposed 
\cite{LKZ92,KLZ92} that their values $\sigma^c_{xy}= (n-1/2)\,e^2/h$ 
and $\sigma^c_{xx}=1/2\,e^2/h$ are simply related at the respective
$n$-th transition.

Away from the critical points, the Hall conductivity is exactly quantized 
due to topological reasons. Thus, in the presence of electron-hole symmetry, 
the above formula for $\sigma_{xy}$ is evident at the transition points 
\cite{HHB93}. Support for the existence of a universal $\sigma_{xx}$ comes both 
from experiments \cite{WJTP95, Sea95b} and numerical calculations. Amongst 
the latter, a critical conductivity $\sigma^c_{xx}=0.54\,e^2/h$ 
was found for the lowest Landau band using white noise disorder potentials 
\cite{CD88}. A similar diagonalization study \cite{HHB93} 
yielded $\sigma^c_{xx}=(0.55\pm 0.05)\,e^2/h$ for various correlated random 
potentials, a result considered by the authors to be consistent with the
prediction \cite{LKZ92} $\sigma^c_{xx}=1/2\,e^2/h$. However, the critical 
diagonal conductivity deduced from a Thouless number study \cite{And89a}
exhibited a strong dependence on the range of the scatterers.
Moreover, calculations of the two-terminal conductance of square systems 
produced a significantly larger value $g_c=(0.58\pm\,0.03)\,e^2/h$ \cite{WJL96}
and showed considerable fluctuations exhibiting a broad distribution between 
zero and $e^2/h$ \cite{CF97}. Here $g_c=\lim_{L\to\infty}\sum_i g_i^{c}(L)/N$ 
is the disorder averaged critical conductance of $N$ realizations. 
This value was derived
from a finite-size scaling study of the transfer-matrix results calculated 
within a Chalker-Coddington (CC) network model \cite{CC88}. In a recent work, 
this outcome has been confirmed with $g_c=(0.57\pm\,0.02)\,e^2/h$ \cite{OSK04}.  

On the other hand, similar transfer-matrix calculations within a tight-binding 
(TB) lattice model with spatially uncorrelated disorder potentials yielded 
$g_c=(0.506\pm\,0.01)\,e^2/h$ \cite{WLS98}, hence, the asserted universality
cannot be taken for granted. 
Even if it were possible to reconcile the contradicting results, the hitherto 
unknown relation between the two-terminal conductance and the conductivity, 
the latter calculated for instance from the Kubo formula for systems without
contacts and  leads, would still inhibit us drawing final conclusions about the 
validity of the proposed universal value of the critical dissipative 
conductivity $\sigma^c_{xx}=1/2\,e^2/h$ \cite{LKZ92}. 

Our work is aimed at resolving this issue by making available the missing 
links and therewith provide evidence for an equality of conductivity and 
conductance at the critical point in the lowest Landau band. We find, within
the  uncertainty of our calculations, similar values $0.58\,e^2/h$
and $0.6\,e^2/h$, respectively. 
These values are, however, significantly larger than 
the previously proposed and well accepted $\sigma^c_{xx}=1/2\,e^2/h$. 
The origin of this discrepancy is suggested to arise from the 
multifractal correlations of the critical eigenstate amplitudes, 
which were presumably not fully taken into account previously. 
We also present the critical conductance distribution for the second
Landau band which agrees with the one from the lowest. The scaling of the
ensemble averaged $g_c$ yields $(0.61\pm 0.03)\,e^2/h$, a result that supplies
further evidence for the existence of a universal conductance.  

We first show that for non-interacting electrons moving in a quenched
disorder potential and perpendicular magnetic field described by a 
tight-binding lattice model the following relation holds,  
\begin{equation}
g_c(L)=g_c(\infty) - g_0 \, (L_0/L)^{y},
\end{equation} 
if the size of the square systems is $L>L_0$. We find in the lowest Landau
band a critical conductance $g_c(\infty)=(0.60\pm 0.02)\,e^2/h$ with 
$y=0.4\pm 0.02$. 
Such a relation has been suggested previously \cite{Pol98} where the least 
irrelevant scaling exponent $y$ was proposed to equal the exponent 
of the critical eigenfunction correlations \cite{CD88}. 
Then, we demonstrate that the Kubo-conductivity for 
very long strips of width $M>M_0$ obeys a similar power law 
\begin{equation}
\sigma^c_{xx}(M)=\sigma^c_{xx}(\infty)  - \sigma_0 \, (M_0/M)^{y_1},
\label{CritC}
\end{equation} 
with $\sigma^c_{xx}(\infty)=(0.58\pm 0.02)\,e^2/h$. The irrelevant scaling 
exponent $y_1$ is $0.38\pm 0.03$. Our value of the critical conductivity is 
in good agreement with the two-terminal conductance, 
but is significantly larger than predicted
previously \cite{LKZ92}.

The movement of non-interacting charge carriers in a disordered
two-dimensional system in the presence of a perpendicular magnetic field
$B=\phi \hbar/(ea^2)$ is described by a tight-binding  model on a square
lattice with lattice constant $a$, and $\phi/(2\pi)$ is the number of flux 
quanta per plaquette,  
\begin{eqnarray}
{\cal H}/V&=&\sum_{xy}w_{xy}c^{\dagger}_{xy}c^{}_{xy}
+c^{\dagger}_{x y+a}c^{}_{xy}+c^{\dagger}_{x y-a}c^{}_{xy}\\\nonumber
&+& \sum_{xy}\exp(i\phi y)\,c^{\dagger}_{x+a y}c^{}_{xy}
+\exp(-i\phi y)\,c^{\dagger}_{x-a y}c^{}_{xy}.
\end{eqnarray}
The distribution of spatially Gaussian correlated on-site disorder potentials 
$w_{xy}$ with zero mean has been generated as described in
Ref.~\onlinecite{KS03} from uncorrelated random numbers 
evenly distributed between $[-W/V,W/V]$. Here, 
$W$ is the disorder strength in units of the transfer term $V$ and 
$\langle w_{xy}w_{x'y'}\rangle \sim
\exp(-[(x-x')^2+(y-y')^2]/(2\kappa^2))$ defines the potential's correlation length
$l_p=\sqrt{2}\kappa$ which was varied in the range $0 < \kappa \lesssim l_B$, 
where $l_B=a/\sqrt{\phi}$ is the magnetic length. 

In contrast to the one-band model \cite{CD88,HHB93} or the network model 
\cite{WJL96,OSK04}, the energetical position of the transition points 
are usually not known \textit{a priori} in the TB model. They depend on 
disorder strength, correlation length, and magnetic field. Therefore, careful 
extensive calculations had to be carried out to precisely detect the critical 
points. A slight imprecision influences considerably the results for the 
conductance and the conductivity.  We used various methods like
calculations of the energy and size dependence of level statistics
\cite{KS03}, localization length \cite{KPS01}, and Hall conductivity to find
the exact critical energy. The first quantity was obtained by direct
diagonalization whereas the latter two were calculated by means of a recursive
Green function technique \cite{Mac85}.
The two-terminal conductance of a $L\times L$ sample with two semi-infinite
ideal leads attached to opposite sides and periodic boundary conditions
applied in the transverse direction was calculated via the Landauer formula
$g=\mathrm{Tr}\,t^{\dagger}t$ where $t$ is the transmission matrix. For a given
disorder realization the conductance was numerically evaluated using a well known
algorithm \cite{PMR92}. Ensembles of at least $N=10^5$ realizations were used 
($N>10^4$ in case of $L/a=512$) to determine the conductance distribution
and the corresponding mean critical conductance $g_c(L)$.   
The critical conductivity was obtained within linear response theory using a 
recursive Green function method \cite{Mac85}.

\begin{figure}
\includegraphics[width=7.5cm]{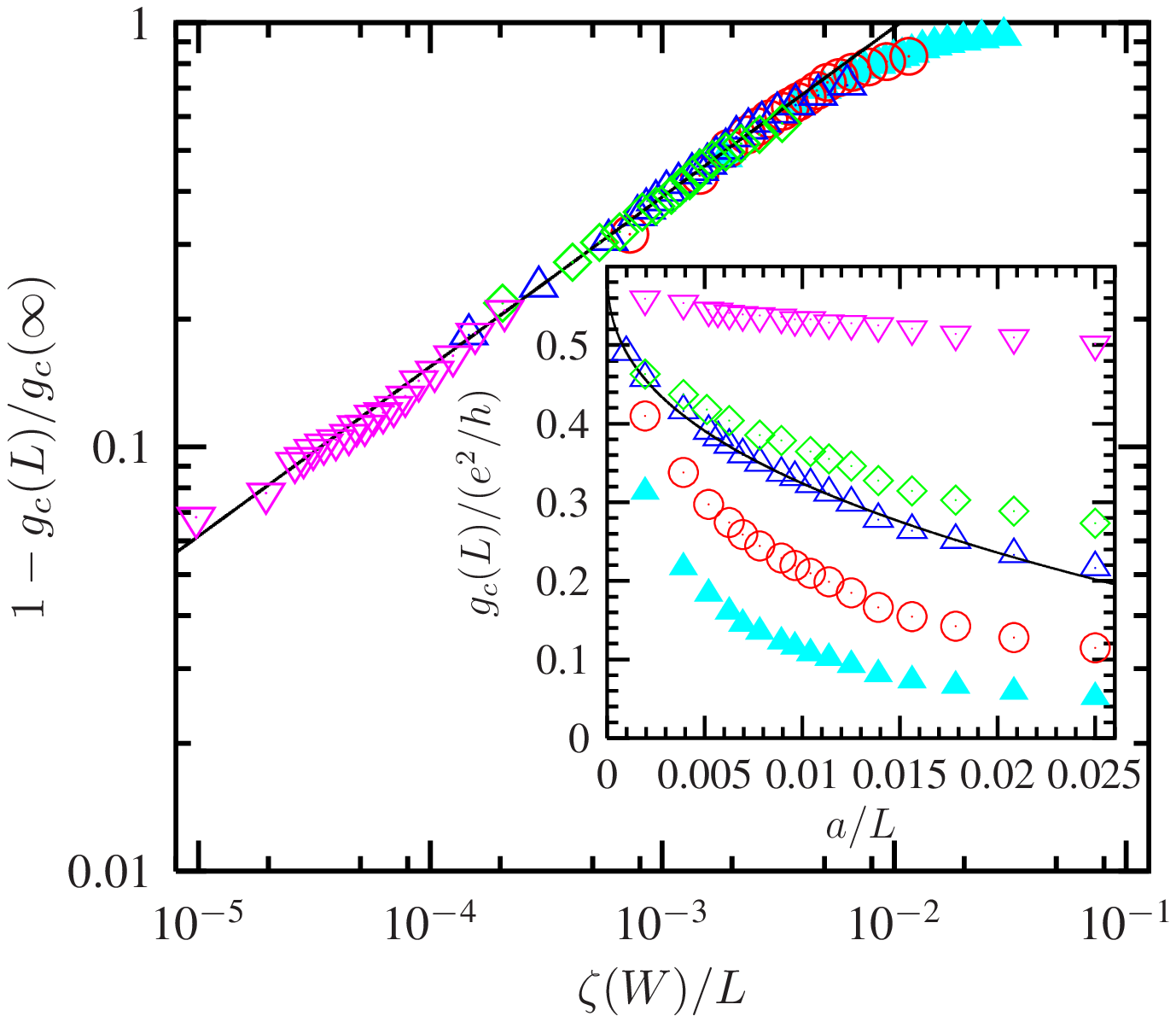}
\caption{(color online). The relative deviation of the mean critical conductance 
  $g_c(L)$ from $g_c(\infty)$ vs.\@ rescaled inverse system size $L$ for a 
  disorder potential with correlation $\kappa/a=1.0$, disorder strengths
  $W/V=0.125$ (\UParrow), $0.25$ (\Circle), $0.5$ (\APLup), $0.7$ ($\Diamond$),
  $6.0$ (\APLdown), and $B=1/(8a^2)\cdot h/e$. The best fit 
  gives $g_c(\infty)=0.6\pm 0.02$ and a slope $y=0.4$.
  The inset shows the unscaled data for $L/a > 40$.}
\label{delg_1stB8e1p0_gLW}
\end{figure}

The mean critical conductance $g_c$
depends on the system's size $L$, the disorder strength $W$, and the
correlation length $l_p=\sqrt{2}\kappa$ of the spatially correlated disorder 
potentials. 
In order to detect the genuine size dependence and to be able to correlate 
our results with those obtained in the presence of only one Landau band
\cite{CD88,HHB93,WJL96,OSK04}, one has to confine the parameter range to the 
situation where the width of the disorder broadened Landau bands
$\Gamma(W,\kappa)$ is small compared to their energy separation $\hbar\omega_c$.   
Stronger disorder enhances those matrix elements that couple neighboring
Landau bands giving rise to an increasing $g_c$ due to extra scattering events
into the next band, and eventually, shifts the critical states to higher energy
until they disappear. This levitation of the current carrying states was shown
to depend essentially on the correlation length \cite{KPS01}.
The size dependence of the mean conductance was calculated for various
disorder strengths $W$ and correlation parameters $\kappa/a=0.3, 1.0$, and $1.5$. 
For disorder potentials with $\kappa/a=1.0$, which corresponds to a correlation
length of the order of the magnetic length ($l_p/l_B=1.2533$), the rescaled mean
conductances are shown in Fig.~\ref{delg_1stB8e1p0_gLW} for system sizes 
$24 \le L/a \le 512$ and disorder strength $0.125 \le W/V \le 0.7$ and
$W/V=6.0$. In addition, 
the average over 1610 realizations for systems of size $L=1024$ and $W=0.5$ 
is also included. The inset shows the original data for $L\ge 40\,a$. 
The best fit, leading to a mean critical conductance $g_c(\infty)=(0.60\pm
0.02)\,e^2/h$ and an irrelevant scaling exponent $y=0.4\pm 0.02$ is also shown
in the inset of Fig.~\ref{delg_1stB8e1p0_gLW} (solid line). Our mean critical 
conductance is in accordance with the results reported for the network model 
\cite{WJL96,OSK04}. 

The rescaling of the data according to $g_c(\infty)-g_c(L) \propto
(\zeta(W)/L)^y$ is in principle possible only for small disorder strength.
For stronger disorder 
($1.0 \le W/V \le 5.0$), the true scaling regime cannot be reached for 
system sizes $L/a < 192$ because Landau band mixing, which is obvious from
the strong overlap seen in the density of states $\varrho$, causes scattering
into the critical states in higher Landau bands. 
This disorder induced Landau band coupling may be the reason for the
differing result $g_c=(0.506\pm\,0.01)\,e^2/h$ of Ref.~\onlinecite{WLS98}.
Increasing the disorder strength further, one reaches the situation where
critical states in higher Landau bands get annihilated by their corresponding
anti-Chern states \cite{KS03}. For $W/V=6.0$, when only the critical state of
the lowest Landau band survives, scattering into higher conducting states is
not possible so that the conductance data again show scaling and match
up with those obtained for small disorder. 
A similar behavior is observed for weak correlated disorder potentials with
$\kappa/a=0.3$, where $l_p=0.424\,a$ is less than half the magnetic length.
In this case, we found $g_c(\infty)=0.59\pm 0.03$ and $y=0.42\pm 0.03$ from the
rescaled data (not shown). 

\begin{figure}
\includegraphics[width=7.0cm]{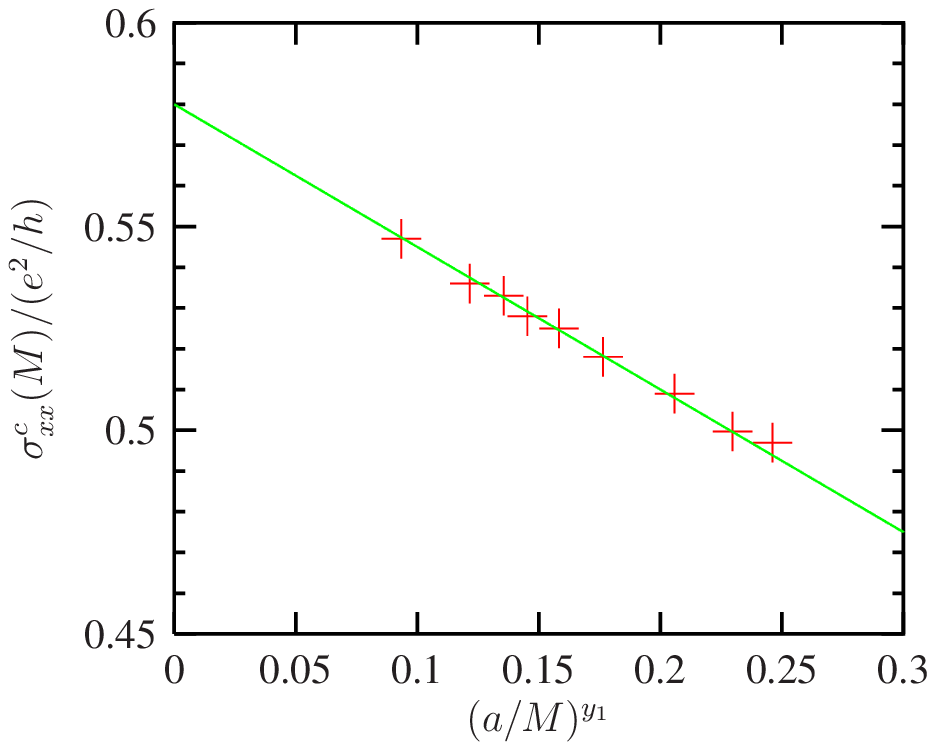}
\caption{(color online). The critical longitudinal conductivity vs.\@
  inverse system width $M$ for a correlated disorder potential with $W=0.5$, 
  $\kappa/a=0.3$, and magnetic flux density $B=1/(8a^2)\cdot h/e$. The fit to   
  Eq.~(\ref{CritC}) gives $\sigma^c_{xx}(\infty)=(0.58\pm 0.02)\,e^2/h$ and
  $y_1=0.38\pm 0.03$.}  
\label{sxxc_e0p3_irr0p38}
\end{figure}

Having established the critical conductance in the lowest Landau band, we turn  
now to our results obtained from the investigation of the critical Kubo
conductivity for disorder potentials with $\kappa/a=0.3$. In
Fig.~\ref{sxxc_e0p3_irr0p38} the dependence of $\sigma^c_{xx}$ 
on the width of the system is shown in the range $40\le M/a\le 512$.
We find in the large size limit $\sigma^c_{xx}=(0.58\pm 0.03)\,e^2/h$ which is 
significantly larger than proposed previously. 

The reason for a critical conductance larger than $0.5\,e^2/h$ and,
therefore, for the discrepancy with the suggested behavior involving
quantum percolation at the critical point in the integer QHE 
\cite{KLZ92,LKZ92,LWK93} 
may be associated with the multifractality of the critical quantum Hall 
eigenstates \cite{Aok86,HS92,PJ91,HKS92,HS94}, a property that was 
not fully appreciated at that time.  
The diffusion coefficient $D$, determining the conductivity
$\sigma_{xx}=e^2\varrho D$, is essentially influenced by the multifractal
correlations of the wave function amplitudes \cite{Cha90,CD88,HS94,BHS96}. 
Fal'ko and Efetov have shown \cite{FE95,FE95a} that for length scales shorter 
that the localization length, these correlations are also present away from the
Anderson transition. Using a special version of the supersymmetric $\sigma$
model, they provided a formula relating the diffusion coefficient
to the generalized multifractal dimensions $d(q)$, 
\begin{figure}
\includegraphics[width=8.cm]{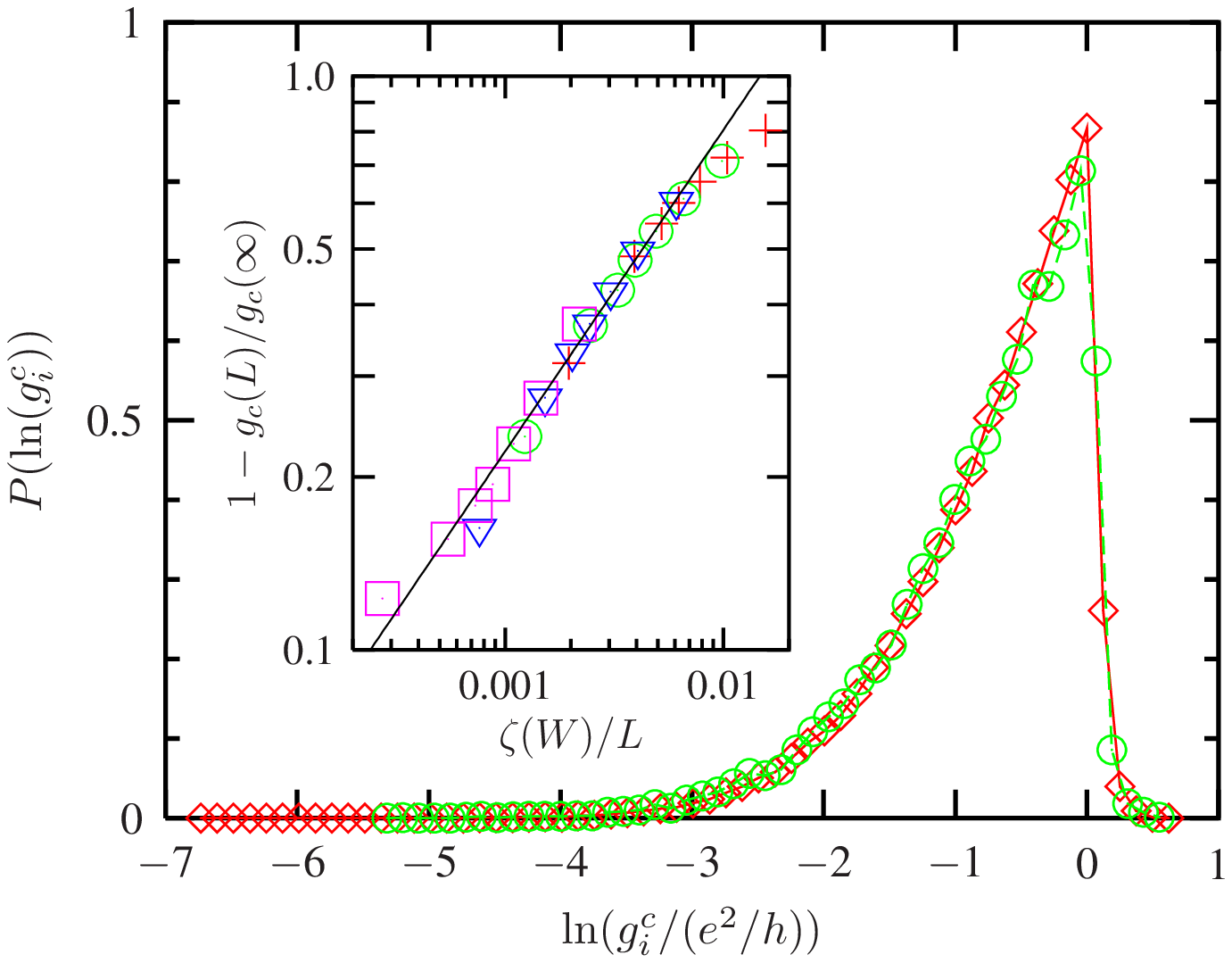}
\caption{(color online). The critical conductance distribution
  functions $P(\ln(g_i^{c}))$ from the lowest ({\large $\diamond$}, $L/a=192,
  W/V=6.0, \kappa/a=1.0, B=1/(8a^2)\cdot h/e$) and the  second lowest ({\large
  $\circ$}, $L/a=512, W/V=0.5, \kappa/a=1.5, B=1/(32a^2)\cdot h/e$) Landau
  band (LB). 
  The inset shows the relative deviation of the mean critical conductance $g_c(L$) 
  for the 2nd LB vs.\@ rescaled inverse system size $L$ for a disorder potential 
  with correlation parameter $\kappa/a=1.5$ and disorder strengths $W/V=0.125$ ($+$), 
  $0.18$ (\Circle), $0.25$ (\APLdown), and $0.5$ ($\Box$). The best fit gives
  $g_c(\infty)=(0.61\pm 0.03)\,e^2/h$ and a slope $y=0.56\pm 0.05$}
\label{2ndLB}
\end{figure}
\begin{equation}
d(q) = 2-\frac{q}{4\pi^2\varrho D \hbar}.
\label{FalEfe}
\end{equation}
This relation for unitary symmetry was obtained in the leading order in
$1/(2\pi \varrho D \hbar)$ and should hold for $q\le 2\pi \varrho D \hbar$. 
We assume Eq.~(\ref{FalEfe}) to be exact near the quantum Hall critical
point for electronic states with localization length larger than the
system size \cite{FE95a}.  Taking $q=1$, we find very good agreement with our  
numerical results presented above, 
\begin{equation}
\sigma_{xx}=\frac{1}{2\pi(2-d(1))}\,\frac{e^2}{h}=0.61\,\frac{e^2}{h},
\label{critcond}
\end{equation}
using the information dimension $d(1)=\alpha(1)=1.739\pm 0.002$ derived
\cite{EMM01} from the precise parabolic $f(\alpha(q))$ singularity strength
distribution. Over the years, convincing numerical evidence for a universal 
$f(\alpha(q))=2-(\alpha(q)-\alpha_0)^2/[4(\alpha_0-2)]$ has been accumulated
for different quantum Hall models \cite{PJ91,HKS92,KM95,KM99,EMM01}.  Within
error 
bars, the whole distribution appears to be independent of disorder strength, 
potential correlation length, and magnetic field. It is determined only by a
single value $\alpha_0=2.262\pm0.003$ \cite{EMM01}. 
Therefore, also $d(1)$ must be universal and so should the critical
conductivity, if Eq.~(\ref{critcond}) was exact.

We also found the multifractal properties of the wavefunctions to be
independent of the Landau band index for correlated disorder in agreement 
with previous investigations \cite{TNA96}.
Therefore, to check the universality of Eq.~(\ref{critcond}), we calculated 
the critical conductance in the 2nd Landau band within the same model, but for 
weaker magnetic field $B=1/(32a^2)\cdot h/e$ and stronger disorder
correlations $\kappa/a=1.5$ ($l_p/l_B=0.94$). 
We found $g_c(\infty)= (0.61\pm 0.03)\,e^2/h$ (see inset of Fig.~\ref{2ndLB}) 
from the scaling of the mean critical two-terminal conductance, which is in
good agreement with our results for the lowest Landau band.
The critical conductance distribution functions $P(g_{i}^{c}(L))$ for the 1st
and 2nd Landau band of our TB model (shown in Fig.~\ref{2ndLB}) are almost
identical and look the same as the one reported for the CC model \cite{JW98}. 
Conductance values larger that $e^2/h$ reflect the fact that more than one
transmission channels are effective, due to the two-dimensional contacts.
This is contrary to investigations of the critical point-contact conductance 
\cite{JMZ99,CRSR01} which show a flat symmetrical distribution between
zero and $e^2/h$ and, therefore, a mean conductance $g_c=0.5\,e^2/h$.  

A universal critical conductivity $\sigma_{xx}^c\simeq 0.6\,e^2/h$
should be easily observable in experiments on Corbino disks in the low
temperature limit. Here, the disagreement with the value $1/2\,e^2/h$
should be clearly discernible and, for the first time, reveal the 
multifractality of the critical eigenstates in a transport measurement. 
In case of the usual Hall bars, using our new value for the
critical conductivity, the critical resistance 
$\rho_{xx}^c=\sigma_{xx}^c/[(\sigma_{xx}^c)^2 + (\sigma_{xy}^c)^2]$  
turns out to be only a little smaller, $0.98\,h/e^2$ instead of
$\rho_{xx}^c=1\,h/e^2$, for 
the transition point in the lowest Landau band, but with $0.23\,h/e^2$
somewhat larger than $\rho_{xx}^c=0.2\,h/e^2$ at the second transition. 
As a further test of the suggested general influence of the eigenfunction
correlations on the critical conductance, we investigated also a
two-dimensional disordered system with spin-orbit interactions. 
Using the Ando model, we obtained from transfer matrix calculations a critical
conductance $g^s_c=(1.43\pm 0.05)\,e^2/h$ and, via direct diagonalization and 
a multifractal analysis of the critical eigenfunctions, 
an information dimension $d^s(1)=1.89$. We find again that both
quantities are perfectly related by a symplectic version of Eq.~(\ref{FalEfe})
\cite{FE95a}. 

In conclusion, we have shown that in the critical quantum Hall regime both the
averaged conductance of square samples and the conductivity of  very long
strips converge to the same universal value $\simeq 0.6\,e^2/h$
in the limit of large system size.  This value does neither depend on the
considered disorder realizations nor on the Landau level index, and it is
consistent with theoretical predictions that take in to account
multifractal eigenstate correlations at quantum critical points.

PM thanks VEGA, grant No.\@ 2/3108/2003, for financial support and PTB for its
hospitality.


\end{document}